\UseRawInputEncoding
\documentclass[preprint,12pt]{elsarticle}%
\usepackage{amssymb}
\usepackage{amsfonts}
\usepackage{amsmath}
\usepackage{graphicx}%
\providecommand{\U}[1]{\protect\rule{.1in}{.1in}}
\graphicspath{{E:/dottorato/Coulomb_damper/Testo/Paper/Immagini/}}
\journal{journal}

\begin{document}
%
%TCIMACRO{\TeXButton{Begin frontmatter}{\begin{frontmatter}}}%
%BeginExpansion
\begin{frontmatter}%
%EndExpansion



%TCIMACRO{\TeXButton{Title}{\title
%{Crack propagation at the interface between viscoelastic and elastic materials}%
%}}%
%BeginExpansion
\title
{Crack propagation at the interface between viscoelastic and elastic materials}%
%EndExpansion


%TCIMACRO{\TeXButton{Author}{\author{M. Ciavarella(1,2), R.McMeeking(3,4,5,6)}%
%}}%
%BeginExpansion
\author{M. Ciavarella(1,2), R.McMeeking(3,4,5,6)}%
%EndExpansion
%

%TCIMACRO{\TeXButton{Address}{\address
%{(1) Politecnico di BARI. DMMM department. Viale Gentile 182, 70126 Bari. Mciava@poliba.it,
%(2) Hamburg University of Technology, Department of Mechanical Engineering, Am Schwarzenberg-Campus 1, 21073 Hamburg, Germany
%(3) Materials Department, University of California, Santa Barbara CA 93106
%(4) Department of Mechanical Engineering, University of California, Santa Barbara CA 93106
%(5) School of Engineering, Aberdeen University, King's College, Aberdeen AB24 3UE, Scotland
%(6) INM - Leibniz Institute for New Materials, Campus D2 2, 66123 Saarbrücken, Germany}%
%}}%
%BeginExpansion
\address
{(1) Politecnico di BARI. DMMM department. Viale Gentile 182, 70126 Bari. Mciava@poliba.it,
(2) Hamburg University of Technology, Department of Mechanical Engineering, Am Schwarzenberg-Campus 1, 21073 Hamburg, Germany
(3) Materials Department, University of California, Santa Barbara CA 93106
(4) Department of Mechanical Engineering, University of California, Santa Barbara CA 93106
(5) School of Engineering, Aberdeen University, King's College, Aberdeen AB24 3UE, Scotland
(6) INM - Leibniz Institute for New Materials, Campus D2 2, 66123 Saarbrücken, Germany}%
%EndExpansion
%

%TCIMACRO{\TeXButton{Begin abstract}{\begin{abstract}}}%
%BeginExpansion
\begin{abstract}%
%EndExpansion

Crack propagation in viscoelastic materials has been understood with the use
of Barenblatt cohesive models by many authors since the 1970's. In polymers
and metal creep, it is customary to assume that the relaxed modulus is zero,
so that we have typically a crack speed which depends on some power of the
stress intensity factor. Generally, when there is a finite relaxed modulus, it
has been shown that the toughness increases between a value at very low speeds
at a threshold toughness $G_{0}$, to a very fast fracture value at $G_{\infty
}$, and that the enhancement factor in infinite systems (where the classical
singular fracture mechanics field dominates) simply corresponds to the ratio
of instantaneous to relaxed elastic moduli.

Here, we apply a cohesive model for the case of a bimaterial interface between
an elastic and a viscoelastic material, assuming the crack remains at the
interface, and neglect the details of bimaterial singularity. For the case of
a Maxwell material at low speeds the crack propagates with a speed which
depends only on viscosity, and the fourth power of the stress intensity
factor, and not on the elastic moduli of either material. For the\ Schapery
type of power law material with no relaxation modulus, there are more general
results. For arbitrary viscoelastic materials with nonzero relaxed modulus, we
show that the maximum toughness enhancement will be reduced with respect to
that of a classical viscoelastic crack in homogeneous material.%

%TCIMACRO{\TeXButton{End abstract}{\end{abstract}}}%
%BeginExpansion
\end{abstract}%
%EndExpansion
%

%TCIMACRO{\TeXButton{Begin keyword(s)}{\begin{keyword}}}%
%BeginExpansion
\begin{keyword}%
%EndExpansion

Viscoelasticity, crack propagation, cohesive models, bimaterial interfaces%

%TCIMACRO{\TeXButton{End keyword(s)}{\end{keyword}}}%
%BeginExpansion
\end{keyword}%
%EndExpansion
%

%TCIMACRO{\TeXButton{End frontmatter}{\end{frontmatter}}}%
%BeginExpansion
\end{frontmatter}%
%EndExpansion



\section{\bigskip Introduction}

The problem of viscoelastic crack growth is of fundamental importance see the
recent review of Rodriguez et al. (2020). Early investigations (Gent and
Schultz, 1972, Barquins and Maugis 1981, Gent, 1996, Gent \& Petrich 1969,
Andrews \& Kinloch, 1974, Barber \textit{et al}, 1989, Greenwood \& Johnson,
1981, Maugis \& Barquins, 1980) noticed a steady state subcritical crack
propagation with an enhanced work of adhesion $G$ with respect to the
adiabatic value $G_{0}$ namely $\ $%
\begin{equation}
\frac{G}{G_{0}}=1+\left(  \frac{V}{V_{0}}\right)  ^{n} \label{wvisco}%
\end{equation}
where $V$ is velocity of peeling of the contact/crack line, a characteristic
velocity was defined as $V_{0}=\left(  ka_{T}^{n}\right)  ^{-1}$ and $k,n$ are
(supposedly) constants of the material, with $0<n<1$ where $a_{T}$ is the WLF
factor to translate results at various temperatures $T$ (Williams, Landel \&
Ferry, 1955).

From a more fundamental perspective, initially a paradox was identified by
Graham (1969). Namely, since the stress field singularity is the classical
inverse square root of elastic materials, at the crack tip we have infinite
frequency and therefore an "elastic" material, which does not explain
dissipation and speed dependence of toughness enhancement (Rice, 1978). But
the paradox was solved by various authors (Schapery, 1975, Greenwood
\&\ Johnson, 1981, Barber \textit{et al.}, 1989, Greenwood, 2004 and others,
see a review by Bradley et al.1997) using cohesive Barenblatt or Dugdale
models removing the singularity in a cohesive zone whose size increases with
speed (because of the toughness enhancement). Another school explains
enhancement with estimating the dissipation (de Gennes, 1996), by the
"viscoelastic trumpet" crack model, as the crack shape is different in the
inner "glassy region", the intermediate "liquid region", and finally in the
outer soft "rubbery" region. On this second school, notable improvements have
been made by Persson \&\ Brener (2005) who clarified the relevant range of
frequencies involved in the estimating dissipation, and gave a solution for a
general viscoelastic material.

Both schools suggest results qualitatively of the form (\ref{wvisco}) and
introduce the cohesive strength of the material $\sigma_{c}$ and therefore
introduce the length scale
\begin{equation}
a_{0}=\frac{G_{0}E_{0}}{2\pi\sigma_{c}^{2}} \label{a0}%
\end{equation}
where $E_{0}$ is the relaxed modulus (the modulus at zero frequency) and
$\sigma_{c}$ is the cohesive stress.

Also, all these models suggest looking at the remote stress intensity factor
$K$ as applied in remote regions as giving an effective toughness $K_{Ic}%
^{2}\left(  V\right)  =G\left(  V\right)  E_{0}$, where $E_{0}$ is expected in
remote regions, and hence obtain the maximum toughness enhancement as
\begin{equation}
\frac{G\left(  \infty\right)  }{G_{0}}=\frac{E_{\infty}}{E_{0}}%
\end{equation}

For many polymeric or rubbery materials this ratio is very large possibly an
increase of 3 or 4 orders of magnitude. Here, $G\left(  \infty\right)  $
stands for $G\left(  V\rightarrow\infty\right)  $ and $E_{\infty}$ is
instantaneous modulus or the modulus observed at infinite frequency of
oscillatory loading. Remark however that in much of the literature on metal
creep or polymers like that reviewed in Bradley et al.(1997), it is often
assumed that the rheology corresponds to a zero relaxed modulus $E_{0}=0$ and
hence there is no lower threshold for crack propagation. These studies obtain
a power law for the crack propagation speed, of the type%
\begin{equation}
V\propto K_{I}^{m}%
\end{equation}
where $m$ depends on details of the rheology. Notice that the Gent-Schultz
kind of result Eq. (\ref{wvisco}) is fully compatible with this scaling, when
one considers $G>>G_{0}$ so that (\ref{wvisco}) can be written as $V\propto
K_{I}^{2/n}$ i.e. $m=2/n$. For a standard material it can be shown that in the
intermediate range of velocities $n=1/2$ and hence $V\propto K_{I}^{4}$.

\bigskip Therefore, as clearly explained by Wang et al (2016), the literature
on ceramics or metal creep and on non cross-linked polymers see
viscoelasticity as an "apparent weakening", since they compare to the elastic
fast fracture limit, while the polymers literature shows viscoelastic results
as a mechanism for enhanced dissipation "an apparent toughening" with respect
to the threshold for the start of subcritical crack growth.

However, analysis of the case of a bimaterial crack between an elastic
material and a viscoelastic one has not been attempted before to the best of
our knowledge, at least not from a simplified perspective as we shall provide
here. This is surprising, since the bimaterial crack problem has received much
attention for the elastic-elastic case (there are at least two papers with
more than 1000 citations, for example, Suo \&\ Hutchinson, 1990, and Comninou,
1977), while in many cases interface cracks occur between a rubbery or
polymeric material peeling from an elastic surface (see Kendall's classical
paper, Kendall (1975) again with nearly 1000 citations). This topic is now an
emerging area of technology in adhesives and in Nature-inspired attachment
systems, and in many cases the polymer will exhibit viscoelasticity. Therefore
an analysis of the the problem is timely. Indeed, we show here a simple
generalization of the cohesive model treatment for a bimaterial interface with
viscoelasticity, obtaining some (approximate) closed form results.

\section{A bimaterial crack propagation}

If we consider a bimaterial interface with a semi-infinite crack, see Fig.1,
where one material is elastic and the other viscoelastic, we can assume a
cohesive model and write the Energy Release Rate (ERR) as
\begin{equation}
G=G_{e}+G_{v}=G_{0}\label{balance}%
\end{equation}
where $G_{e}$ is the ERR in the elastic material, and $G_{v}$ the ERR\ in the
viscoelastic material.

\begin{center}%
\begin{tabular}
[c]{l}%
\centering\includegraphics[height=65mm]{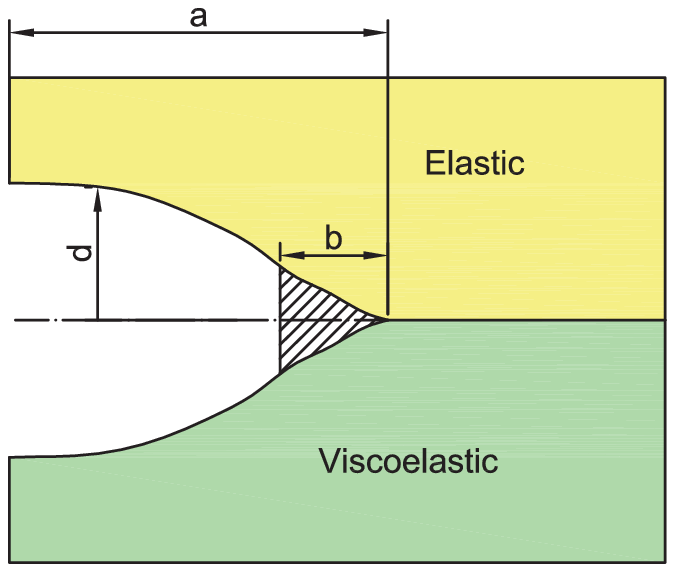}
\end{tabular}

Fig.1 - The crack of size $a>>b$ where $b$ is a cohesive region, propagating
at the interface between an elastic and a viscoelastic material
\end{center}

We consider the classical Schapery-Greenwood Dugdale cohesive model, and to
address a semi-infinite crack, so we assume the conditions for K-field
dominance, or Small Scale Cohesion (SSC) to be satisfied. In this case we have
simply
\begin{equation}
G_{e}=\frac{1}{2}\frac{K_{I}^{2}}{E}\qquad;\qquad G_{v}=\frac{1}{2}K_{I}%
^{2}C_{eff}\left(  \frac{b}{V}\right)  \label{G}%
\end{equation}
where we consider plane stress for simplicity. For the viscoelastic material,
we have introduced an effective compliance $C_{eff}$ given by (Rice, 1978)
\begin{equation}
C_{eff}\left(  \frac{b}{V}\right)  =\int_{0}^{1}C\left(  \frac{b}{V}-\frac
{b}{V}\lambda\right)  \frac{df}{d\lambda}d\lambda\label{integral}%
\end{equation}
where $C\left(  t\right)  $ is the viscoelastic compliance, relating the
instantaneous strain $\varepsilon\left(  t\right)  =\int_{-\infty}^{t}C\left(
t-\tau\right)  \frac{d\sigma\left(  \tau\right)  }{dt}d\tau$ to the history of
stress $\sigma\left(  t\right)  $ in uniaxial conditions. The function
$f\left(  \lambda\right)  $ is the opening (stretch) in the cohesive zone from
the Dugdale model, given approximately in a bimaterial by%
\begin{equation}
f\left(  \lambda\right)  =\sqrt{\lambda}-\frac{1}{2}\left(  1-\lambda\right)
\ln\frac{1+\sqrt{\lambda}}{1-\sqrt{\lambda}}\label{barenblatt}%
\end{equation}
and the cohesive stress $\sigma_{c}$ cancels the singularity (in the Dugdale
form where the cohesive strength is constant and uniform in the cohesive
region). Under SSC (small scale cohesion), the length of this cohesive zone is
approximately
\begin{equation}
b=\frac{\pi}{8}\left(  \frac{K_{I}}{\sigma_{c}}\right)  ^{2}\label{b}%
\end{equation}

In the appendices we justify the use of the Mode I stress intensity factor
instead of the complex stress intensity factor in the rigorous treatment of
bimaterial interfaces, and hence the approximations introduced in the last 3 equations.

\section{A Maxwell material}

We are not interested in giving a full solution for a general viscoelastic
material, as we shall discuss qualitative features later. A convenient case is
however that of a Maxwell material which has no relaxed modulus. In this case
\begin{equation}
C\left(  t\right)  =\frac{1}{E_{\infty}}+\frac{t}{\mu}%
\end{equation}
where $\mu$ is viscosity. A full simple analytical solution is possible then
by performing the integration (\ref{integral})
\begin{equation}
C_{eff}\left(  \frac{b}{V}\right)  =\frac{1}{E_{\infty}}+\frac{b}{3\mu
V}\label{Ceff}%
\end{equation}

Hence, substituting in the energy balance equation (\ref{balance}) (\ref{G},
\ref{Ceff}) we get\footnote{Obviously we could add the elastic halfplane
compliance in the $C\left(  t\right)  $ and $C_{eff}$ function, and then
remove the elastic contribution to $G$, and the results would not change.}
\begin{equation}
G=\frac{1}{2}\frac{K_{I}^{2}}{E}+\frac{1}{2}K_{I}^{2}C_{eff}\left(  \frac
{b}{V}\right)  =\frac{1}{2}K_{I}^{2}\left(  \frac{1}{E}+\frac{1}{E_{\infty}%
}+\frac{b}{3\mu V}\right)  =G_{0}%
\end{equation}
and using (\ref{b}) this reduces to
\begin{equation}
\frac{1}{2}K_{I}^{2}\left(  \frac{1}{E^{\ast}}+\frac{\pi}{24}\left(
\frac{K_{I}}{\sigma_{c}}\right)  ^{2}\frac{1}{\mu V}\right)  =G_{0}
\label{intermediate}%
\end{equation}
where we define an combined modulus
\begin{equation}
\frac{1}{E_{\infty}^{\ast}}=\frac{1}{E}+\frac{1}{E_{\infty}} \label{combined}%
\end{equation}

This (\ref{intermediate}) leads then to the simple solution
\begin{equation}
\mu V=\frac{\pi}{24}\frac{K_{I}^{4}E_{\infty}^{\ast}/\sigma_{c}^{2}}%
{2G_{0}E_{\infty}^{\ast}-K_{I}^{2}} \label{solution}%
\end{equation}

Obviously, the critical condition of fast fracture is at $V\rightarrow\infty$
when (\ref{solution}) gives
\begin{equation}
K_{Ic,\infty}^{2}=2G_{0}E_{\infty}^{\ast} \label{Kcinfinito}%
\end{equation}
Vice versa, for very low $K_{I}^{2}<<2G_{0}E^{\ast}$, we can write from
(\ref{solution})
\begin{equation}
\mu V_{low}=\frac{\pi}{24}\frac{K_{I}^{4}}{2G_{0}\sigma_{c}^{2}}%
\end{equation}
a simple scaling which does \textit{not} depend on $E^{\ast}$, and so on
neither of the elastic moduli of the elastic and viscoelastic materials (but
just on the viscosity). This is a typical results, when the viscous
fracture-length scale is small and the stress field has the classical K-field
dominance, see eqt.25 of Wang et al (2016), where the scaling is common in
small-scale damage-zone models of creep-rupture in linear materials like the
model of Cocks and Ashby (1982).

\section{The Schapery power law forms of creep-compliance}

More general exponents in the power law could be found if one uses a more
general rheology for the material (but still assuming $E_{0}=0$) i.e. with
more than one relaxation times or a continuous spectrum of relaxation times,
like in much of the literature on metal creep or polymers as that reviewed by
Bradley et al.(1997). Indeed, Schapery (1975a,1975b) shows that one can use
\begin{equation}
C_{eff}\left(  t\right)  =\frac{1}{E_{\infty}}+C_{1}\left(  d\frac{b}%
{V}\right)  ^{n}%
\end{equation}
where $d\simeq1/3$ is a corrective factor which depends very weakly on
$n\in\left[  0,1\right]  $ (in our previous Maxwell model we showed it should
be equal exactly to 1/3), and $C_{1}$is a generalized viscosity with
dimensions $[\frac{s^{-n}}{MPa}]$. For our bimaterial interface, repeating the
analysis we obtain
\begin{equation}
G=\frac{1}{2}\frac{K_{I}^{2}}{E}+\frac{1}{2}K_{I}^{2}C_{eff}\left(  \frac
{b}{V}\right)  =\frac{1}{2}K_{I}^{2}\left(  \frac{1}{E}+\frac{1}{E_{\infty}%
}+C_{1}\left(  d\frac{b}{V}\right)  ^{n}\right)  =G_{0}%
\end{equation}
or
\begin{equation}
\frac{1}{2}K_{I}^{2}\left(  \frac{1}{E_{\infty}^{\ast}}+C_{1}\left(  \frac{\pi
d}{8}\right)  ^{n}\left(  \frac{K_{I}}{\sigma_{c}}\right)  ^{2n}\frac{1}%
{V^{n}}\right)  =G_{0}%
\end{equation}
leading to
\begin{equation}
V=\left(  \frac{\pi d}{8}\right)  \frac{\left(  C_{1}E_{\infty}^{\ast}\right)
^{1/n}}{\left(  2E_{\infty}^{\ast}G_{0}-K_{I}^{2}\right)  ^{1/n}}\left(
\frac{K_{I}^{2+2/n}}{\sigma_{c}^{2}}\right)  \label{ris1}%
\end{equation}
which at low load leads to a scaling with $K_{I}^{2+2/n}$
\begin{equation}
V_{low}=\left(  \frac{\pi d}{8}\right)  \frac{\left(  C_{1}\right)  ^{1/n}%
}{\left(  2G_{0}\right)  ^{1/n}}\left(  \frac{K_{I}^{2+2/n}}{\sigma_{c}^{2}%
}\right)
\end{equation}
The last two equations generalize Schapery's result to a crack on a bimaterial
interface using his assumed generalized rheology. Notice that in the power law
regime the role of elastic modulus disappears again for this class of
materials. Notice also that fast fracture occurs at the same level of stress
intensity factor, independently of $n$, as we expect.

We can rewrite (\ref{ris1}) by taking the dimensionless factor $\widehat
{K}^{2}=\frac{K_{I}^{2}}{2E_{\infty}^{\ast}G_{0}}$, and defining $a_{\infty
}^{\ast}=\frac{G_{0}E_{\infty}^{\ast}}{2\pi\sigma_{c}^{2}}$
\begin{equation}
\widehat{V}=\frac{V}{\left(  C_{1}E_{\infty}^{\ast}\right)  ^{1/n}a_{\infty
}^{\ast}}=\frac{\pi^{2}}{6}\frac{\widehat{K}^{2+2/n}}{\left(  1-\widehat
{K}^{2}\right)  ^{1/n}} \label{vadim}%
\end{equation}

Fig.2 shows some example plots of equation (\ref{vadim}).

\begin{center}%
\begin{tabular}
[c]{l}%
\centering\includegraphics[height=65mm]{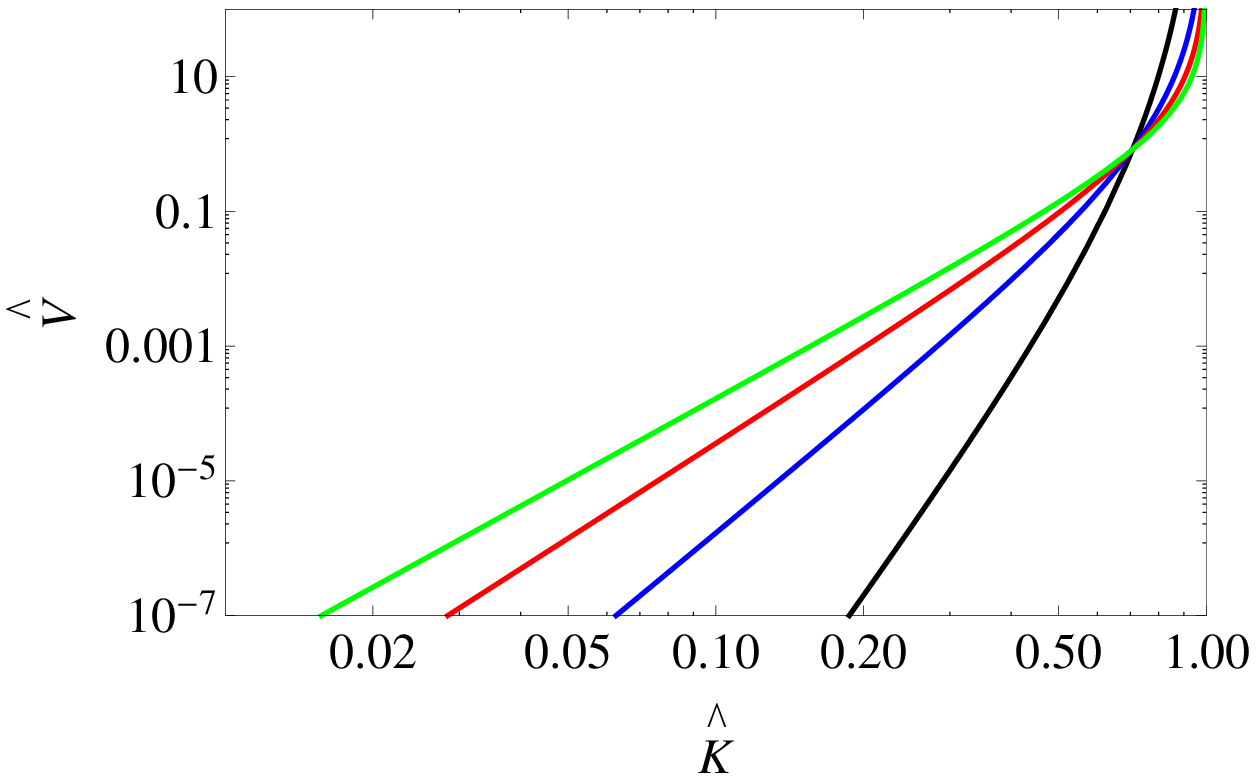}
\end{tabular}

Fig.2 - The dimensionless speed of subcritical crack propagation from equation
(\ref{vadim}) for $n=0.25,0.5,0.75,1$ (black, blue, red, green solid lines) in
a viscoelastic/elastic bimaterial interface for a semi-infinite crack, as a
function of dimensionless stress intensity factor
\end{center}

\section{The case of non-zero relaxed modulus}

Our Maxwell material analytical result is clearly instructive, and the
generalization to the Schapery creep compliance form has proved quite useful
in the engineering literature, see Bradley et al.(1997), as applied to quite
general form of polymers or metal creep (Cocks and Ashby, 1982). It is hard to
give exact simple results for the most general rheology and is outside of the
scope of the present investigation. One point of interest for any viscoelastic
material with finite relaxed modulus $E_{0}>0$, is that obviously
$C_{eff}\left(  V\rightarrow0\right)  =\frac{1}{E_{0}}$, so that we would have
the trivial energy balance equation
\begin{equation}
G=\frac{1}{2}\frac{K_{I}^{2}}{E}+\frac{1}{2}\frac{K_{I}^{2}}{E_{0}}=G_{0}%
\end{equation}
and so we now define another combined modulus
\[
\frac{1}{E_{0}^{\ast}}=\frac{1}{E}+\frac{1}{E_{0}}%
\]
to obtain%
\begin{equation}
K_{Ic,0}^{2}=2G_{0}E_{0}^{\ast}%
\end{equation}

Comparing this with (\ref{Kcinfinito}), we obtain the maximum amplification
\begin{equation}
A_{\max}=\frac{K_{Ic,\infty}^{2}}{K_{Ic,0}^{2}}=\frac{E_{\infty}^{\ast}}%
{E_{0}^{\ast}}=\frac{E_{\infty}}{E_{0}}\frac{\left(  \frac{E_{0}}{E}+1\right)
}{\left(  \frac{E_{\infty}}{E}+1\right)  } \label{amax}%
\end{equation}
and since we expect that for polymers or rubbery materials $\frac{E_{\infty}%
}{E_{0}}>>1$ (which is why we get the toughness amplification)%
\begin{equation}
A_{\max}=\frac{K_{Ic,\infty}^{2}}{K_{Ic,0}^{2}}<\frac{E_{\infty}}{E_{0}}%
\end{equation}

Thus the limiting toughness enhancement is less than that for the crack with
the viscoelastic material on both faces. Fig.3 shows the maximum amplification
$A_{\max}$ from (\ref{amax}) (black thick solid line) as a function of the
ratio $E/E_{0}$, as compared with an even simpler approximation, $A_{\max
}=\frac{E}{E_{0}}$, which shows the maximum amplification is of the same order
of the ratio $E/E_{0}$ in an intermediate range.

\begin{center}%
\begin{tabular}
[c]{l}%
\centering\includegraphics[height=65mm]{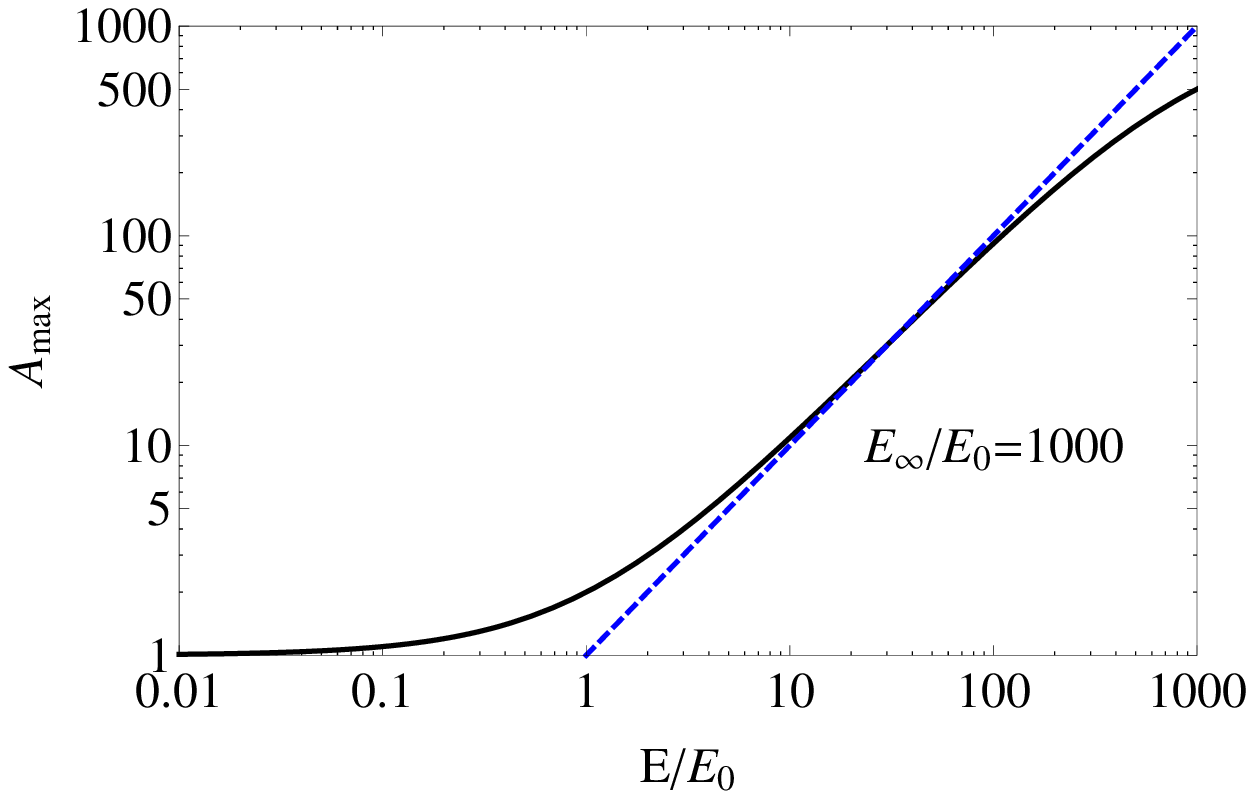}
\end{tabular}

Fig.3 - The maximum toughness amplification $A_{\max}$ from (\ref{amax})
(black thick solid line) as a function of the ratio $E/E_{0}$, where $E$ is
the elastic modulus of the elastic material, and $E_{0}$ the relaxed modulus
of the viscoelastic one. Blue dashed line is the simplified form $A_{\max
}=\frac{E}{E_{0}}$
\end{center}

\section{\bigskip Conclusions}

We have obtained a simple closed form solution for the subcritical propagation
of a crack at the interface between an elastic and a viscoelastic material, in
the form of a Maxwell material, or for the more general creep compliance
rheology of Schapery. We find that the "elastic" fracture occurs for
$K_{Ic,\infty}^{2}=2G_{0}E_{\infty}^{\ast}$ where the "equivalent" modulus is
the inverse of the sum of the compliances of the elastic material and the
(instantaneous value) of the viscoelastic material. The subcritical crack
propagation for the Maxwell material depends only on the viscosity of the
material, and not on any of the elastic moduli, and scales at low speeds with
the fourth order of the stress intensity factor. For the Schapery creep
compliance form, the results are quite similar, except there is a more general
power law dependence. We also argue that for any more general viscoelastic
material constitutive equation, having a relaxed modulus $E_{0}>0$, the
toughness amplification would be less than that expected for the viscoelastic
semi-infinite crack in a homogeneous material.

\section{Acknowledgements}

MC acknowledges support from the Italian Ministry of Education, University and
Research (MIUR) under the program "Departments of Excellence" (L.232/2016). RM
was supported by the MRSEC program of the U.S. National Science Program
through Grant No. DMR-1720256 (IRG3).

\section{References}

Andrews, E. H., \& Kinloch, A. J. (1974). Mechanics of elastomeric adhesion.
In Journal of Polymer Science: Polymer Symposia (Vol. 46, No. 1, pp. 1-14).
New York: Wiley Subscription Services, Inc., A Wiley Company.

Barber, M., Donley, J., \& Langer, J. S. (1989). Steady-state propagation of a
crack in a viscoelastic strip. \textit{Physical Review A}, 40(1), 366.

Barquins, M., \& Maugis, D. (1981). Tackiness of elastomers. \textit{The
Journal of Adhesion}, 13(1), 53-65.).

Bradley, W., Cantwell, W. J., \& Kausch, H. H. (1997). Viscoelastic creep
crack growth: a review of fracture mechanical analyses. Mechanics of
time-dependent materials, 1(3), 241-268.

Burns, S. J., Pollet, J. C., \& Chow, C. L. (1978). Non-linear fracture
mechanics. international Journal of Fracture, 14(3), 311-326.

Cocks ACF, Ashby MF. The growth of a dominant crack in a creeping material.
Scripta Metall 1982;16:109--14.

Comninou, M. (1977). The interface crack. J Appl. Mech. 631-636

de Gennes, P. G. (1996). Soft adhesives. \textit{Langmuir}, 12(19), 4497-4500.

Greenwood J A and Johnson K L (1981) The mechanics of adhesion of viscoelastic
solids \textit{Phil. Mag. A} 43 697--711

Greenwood, J. A. (2004). The theory of viscoelastic crack propagation and
healing. \textit{Journal of Physics D: Applied Physics}, 37(18), 2557.

Gent, A. N., \& Petrich, R. P. (1969). Adhesion of viscoelastic materials to
rigid substrates. \textit{Proceedings of the Royal Society of London. A.
Mathematical and Physical Sciences}, 310(1502), 433-448.

Graham GAC 1969 Two extending crack problems in linear viscoelasticity theory
\textit{Q. Appl. Math.} 27 497--507

Kendall, K. (1975). Thin-film peeling-the elastic term. Journal of Physics D:
Applied Physics, 8(13), 1449.)

Maugis, D., \& Barquins, M. Fracture mechanics and adherence of viscoelastic
solids. In: Adhesion and adsorption of polymers. Springer, Boston, MA, 1980.
p. 203-277.

Persson, B. N. J., \& Brener, E. A. (2005). Crack propagation in viscoelastic
solids. \textit{Physical Review E}, 71(3), 036123.

Rice, J. R. (1978). Mechanics of quasi-static crack growth (No. COO-3084-63;
CONF-780608-3). Brown Univ., Providence, RI (USA). Div. of Engineering.

Rice, J.R., Sih, G.C. (1965) \textquotedblleft Plane problems of cracks in
dissimilar media,\textquotedblright\ \textit{Journal of Applied Mechanics}, 32
pp. 418-423.

Rice, J.R. (1988) \textquotedblleft Elastic fracture mechanics concepts for
interfacial cracks,\textquotedblright\ \textit{Journal of Applied Mechanics}
55 pp. 98-103.

Rodriguez, N., Mangiagalli, P., \& Persson, B. N. J. (2020). Viscoelastic
crack propagation: review of theories and applications. arXiv preprint arXiv:2009.04936.

Schapery R A (1975a) A theory of crack initiation and growth in viscoelastic
media \textit{Int. J. Fracture} 11 (Part I) 141--59

Schapery, R. A. (1975b). A theory of crack initiation and growth in
viscoelastic media II. Approximate methods of analysis. \textit{International
Journal of Fracture}, 11(3), 369-388.

Suo, Z., \& Hutchinson, J. W. (1990). Interface crack between two elastic
layers. International Journal of Fracture, 43(1), 1-18.

Wang, H., Lu, W., Barber, J. R., \& Thouless, M. D. (2016). The roles of
cohesive strength and toughness for crack growth in visco-elastic and creeping
materials. Engineering Fracture Mechanics, 160, 226-237.

Wang, X.-M. , Shen, Y.-P. (1993) \textquotedblleft The Dugdale crack on a
bimaterial interface,\textquotedblright\ \textit{International Journal of
Fracture}, 59 pp. R25-R32.

Williams, M. L.; Landel, R. F.; Ferry, J. D. (1955) The Temperature Dependence
of Relaxation Mechanisms in Amorphous Polymers and Other Glass-Forming
Liquids. \textit{Journal of the American Chemical Society}, 77, 3701-3707.

\section{Appendix}

\noindent The complex stress intensity factor due to a set of point forces
applied to the surface of the crack at $x_{1}={\tilde{x}}_{1}$ is (Rice and
Sih, 1965)

\noindent%
\begin{equation}
K=\sqrt{\frac{2}{\pi}}{\mathrm{cosh}\pi\varepsilon\ }\frac{Q+iR}{{\left(
l-{\tilde{x}}_{1}\right)  }^{\frac{1}{2}+i\varepsilon}} \label{1}%
\end{equation}

\noindent where the $x_{1}$-axis is the bimaterial interface, the crack
occupies $x_{1}\leq l$, where $l=b$ will be the cohesive zone size, and
$\varepsilon$ is the usual bimaterial index%
\begin{equation}
\varepsilon=\frac{1}{2\pi}\ln\frac{1-\beta}{1+\beta}%
\end{equation}
where $\beta$ is Dundurs' bimaterial constant, $i=\sqrt{-1}$, \textit{R}
represents a pair of horizontal, equal and opposite point forces and
\textit{Q} a pair of equal and opposite vertical forces. These forces are
defined per unit thickness. The stress ahead of the crack on the $x_{1}$-axis
is (Rice, 1988)

\noindent%
\begin{equation}
{\sigma}_{22}+i{\sigma}_{12}=\frac{K}{\sqrt{2\pi}}{\left(  x_{1}-l\right)
}^{-\frac{1}{2}+i\varepsilon}%
\end{equation}

\noindent\noindent The stresses in the cohesive zone are uniform and such that
${\sigma}_{22}+i{\sigma}_{12}={\sigma}_{c}+i{\sigma}_{s}$, where ${\sigma}%
_{c}$ is the cohesive stress and ${\sigma}_{s}$ is a shear stress arising as a
reaction to constraints on shearing motion relative to the interface. The net
stress intensity factor is zero, so

\noindent%
\begin{equation}
\sqrt{\frac{2}{\pi}}{\mathrm{cosh}\pi\varepsilon\ }\left(  {\sigma}%
_{c}+i{\sigma}_{s}\right)  \int_{0}^{b}{\frac{d{\tilde{x}}_{1}}{{\left(
b-{\tilde{x}}_{1}\right)  }^{\frac{1}{2}+i\varepsilon}}}=K_{A}%
\end{equation}

\noindent where the cohesive zone occupies $0\leq x_{1}\leq b$, and $K_{A}$ is
the far-field applied stress intensity factor. This integrates to give

\noindent%
\begin{equation}
\sqrt{\frac{8}{\pi}}{\mathrm{cosh}\pi\varepsilon\ }\left(  {\sigma}%
_{c}+i{\sigma}_{s}\right)  \sqrt{b}b^{-i\varepsilon}=\left(  1-2i\varepsilon
\right)  K_{A}%
\end{equation}

\noindent\noindent

We represent the cohesive zone stress as

\noindent%
\begin{equation}
{\sigma}_{c}+i{\sigma}_{s}=\mathrm{\Sigma}e^{i\varphi}%
\end{equation}

\noindent the applied stress intensity factor as

\noindent%
\begin{equation}
K_{A}=\sqrt{K_{A}{\overline{K}}_{A}}e^{i\psi}%
\end{equation}

\noindent and note that

\noindent%
\begin{equation}
b^{-i\varepsilon}=e^{-i\varepsilon{\mathrm{ln}b\ }}%
\end{equation}

\noindent and

\noindent%
\begin{equation}
1-2i\varepsilon={\left(  1+4{\varepsilon}^{2}\right)  }^{\frac{1}{2}%
}e^{-i{\mathrm{arctan}2\varepsilon\ }}%
\end{equation}

\noindent As a result, we find that

\noindent%
\begin{equation}
\varphi=\psi+\varepsilon{\mathrm{ln}b\ }-{\mathrm{arctan}2\varepsilon\ }
\label{9}%
\end{equation}

\noindent%
\begin{equation}
b=\frac{\pi\left(  1+4{\varepsilon}^{2}\right)  K_{A}{\overline{K}}_{A}%
}{8{{\mathrm{cosh}}^{2}\pi\varepsilon\ }T^{2}}%
\end{equation}

\noindent and

\noindent%
\begin{equation}
T^{2}=\frac{{\sigma}_{c}^{2}}{{{\mathrm{cos}}^{2}\left(  \psi+\varepsilon
{\mathrm{ln}b\ }-{\mathrm{arctan}2\varepsilon\ }\right)  \ }} \label{11}%
\end{equation}

\noindent Note that Eq. (\ref{9},\ref{11}) must be solved simultaneously,
probably by iteration.

\noindent\noindent However, for possible material combinations that exclude
auxetics $\varepsilon\ll1$ (Rice, 1988). In that case, a good, 1${}^{st}$
order estimate of $b$ is the result for homogeneous systems

\noindent%
\begin{equation}
b=\frac{\pi K_{A}^{2}}{8{\sigma}_{c}^{2}}%
\end{equation}

\noindent where $K_{A}$ is the mode I stress intensity factor in the far-field
and we have assumed pure Mode I far-field loading.

\noindent

\noindent Now, return to the point force solution Eq.(\ref{1}) and add a
second (auxiliary) set of point forces at $x_{1}={\hat{x}}_{1}$ (notice that
usually ${\hat{x}}_{1}\neq{\tilde{x}}_{1}$) so that the stress intensity
factor is

\noindent

\noindent%
\begin{equation}
K=\sqrt{\frac{2}{\pi}}{\mathrm{cosh}\pi\varepsilon\ }\left[  \frac
{F_{2}+iF_{1}}{{\left(  l-{\tilde{x}}_{1}\right)  }^{\frac{1}{2}+i\varepsilon
}}+\frac{P_{2}+iP_{1}}{{\left(  l-{\hat{x}}_{1}\right)  }^{\frac{1}%
{2}+i\varepsilon}}\right]
\end{equation}

\noindent

\noindent The complex conjugate of \textit{K} is

\noindent%
\begin{equation}
\overline{K}=\sqrt{\frac{2}{\pi}}{\mathrm{cosh}\pi\varepsilon\ }\left[
\frac{F_{2}-iF_{1}}{{\left(  l-{\tilde{x}}_{1}\right)  }^{\frac{1}%
{2}-i\varepsilon}}+\frac{P_{2}-iP_{1}}{{\left(  l-{\hat{x}}_{1}\right)
}^{\frac{1}{2}-i\varepsilon}}\right]
\end{equation}

\noindent

\noindent The resulting energy release rate is (Rice, 1988)

\noindent%
\begin{equation}
G\left(  F_{1},F_{2},P_{1},P_{2},l\right)  =\frac{c_{1}+c_{2}}%
{16{{\mathrm{cosh}}^{2}\pi\varepsilon\ }}K\overline{K} \label{15}%
\end{equation}

\noindent where, in plane strain,

\noindent%
\begin{equation}
c_{1}=\frac{8\left(  1-{\nu}_{1}^{2}\right)  }{E_{1}}%
\end{equation}

\noindent and

\noindent%
\begin{equation}
c_{2}=\frac{8\left(  1-{\nu}_{2}^{2}\right)  }{E_{2}}%
\end{equation}

\noindent with $E_{i}$ being Young's moduli and ${\nu}_{i}$ Poisson ratios.
The subscript 1 indicates the material in $x_{2}\ge0$ and the 2 indicates the
material in $x_{2}\le0$.

\noindent

\noindent Note that both sides of Eq. (\ref{15}) are real. Let the
displacements at $x_{1}={\hat{x}}_{1}$ on the top surface of the crack be
$c_{1}\left(  {\delta}_{1},{\delta}_{2}\right)  /2$, both components defined
to be real. The displacements in the bottom surface of the crack are
$c_{2}\left(  {\delta}_{1},{\delta}_{2}\right)  /2$. Note that

\noindent%
\begin{equation}
{\delta}_{1}={\delta}_{1}\left(  F_{1},F_{2},P_{1},P_{2},l\right)
\end{equation}

\noindent and

\noindent%
\begin{equation}
{\delta}_{2}={\delta}_{2}\left(  F_{1},F_{2},P_{1},P_{2},l\right)
\end{equation}

\bigskip

\noindent Furthermore, using the following generalized Castigliano's theorem

\noindent%
\begin{equation}
{\delta}_{1}=-\frac{\partial\mathrm{\Psi}\left(  F_{1},F_{2},P_{1}%
,P_{2},l\right)  }{\partial P_{1}} \label{delta1}%
\end{equation}

\noindent%
\begin{equation}
{\delta}_{2}=-\frac{\partial\mathrm{\Psi}\left(  F_{1},F_{2},P_{1}%
,P_{2},l\right)  }{\partial P_{2}} \label{delta2}%
\end{equation}

\noindent

\noindent and

\noindent%
\begin{equation}
G=-\frac{\partial\mathrm{\Psi}\left(  F_{1},F_{2},P_{1},P_{2},l\right)
}{\partial l}%
\end{equation}

\noindent where $\mathrm{\Psi}$ is the total potential energy, sum of the
strain energy and the potential energy of the applied loads. Notice that
(\ref{delta1},\ref{delta2}) reduce to the classical Castigliano's theorem for
a linear system, in which $\mathrm{\Psi=-U}$ where $\mathrm{U}$ is strain energy.

\noindent As a result, as noted by Burns et al. (1978) we have Maxwell relationships

\noindent%
\begin{equation}
\left(  \frac{\partial{\delta}_{1}}{\partial l}\right)  _{P_{1}}=\left(
\frac{\partial G}{\partial P_{1}}\right)  _{l}%
\end{equation}

\noindent and

\noindent%
\begin{equation}
\left(  \frac{\partial{\delta}_{2}}{\partial l}\right)  _{P_{2}}=\left(
\frac{\partial G}{\partial P_{2}}\right)  _{l}%
\end{equation}

\noindent Hence, with $P_{1}=P_{2}=0$, these lead to

\noindent%
\begin{equation}
\frac{\partial{\delta}_{2}\left(  {\hat{x}}_{1}\right)  }{\partial l}%
+i\frac{\partial{\delta}_{1}\left(  {\hat{x}}_{1}\right)  }{\partial l}%
=\frac{\left(  c_{1}+c_{2}\right)  \left(  F_{2}+{iF}_{1}\right)  \left(
{\mathrm{cos}\lambda\ }-i{\mathrm{sin}\lambda\ }\right)  }{4\pi{\left(
l-{\tilde{x}}_{1}\right)  }^{\frac{1}{2}}{\left(  l-{\hat{x}}_{1}\right)
}^{\frac{1}{2}}} \label{25}%
\end{equation}

\noindent where
\begin{equation}
\zeta={\mathrm{ln}{\left(  \frac{l-{\tilde{x}}_{1}}{l-{\hat{x}}_{1}}\right)
}^{\varepsilon}\ }%
\end{equation}

\noindent We convert Eq. (\ref{25}) to the result for the cohesive zone with
uniform stress. This leads to

\noindent%
\begin{equation}
\frac{\partial{\delta}_{2}\left(  {\hat{x}}_{1}\right)  }{\partial l}%
+i\frac{\partial{\delta}_{1}\left(  {\hat{x}}_{1}\right)  }{\partial l}%
=-\frac{\left(  c_{1}+c_{2}\right)  \left(  {\sigma}_{c}+{i\sigma}_{s}\right)
}{4\pi{\left(  l-{\hat{x}}_{1}\right)  }^{\frac{1}{2}}}\int_{0}^{l}%
{\frac{\left(  {\mathrm{cos}}\zeta-i{\mathrm{sin}\zeta\ }\right)  d{\tilde{x}%
}_{1}}{{\left(  l-{\tilde{x}}_{1}\right)  }^{\frac{1}{2}}}}%
\end{equation}

\noindent With $\varepsilon$ small,
\begin{equation}
{\mathrm{cos}\zeta\simeq1-}\frac{\zeta^{2}}{2}={1-\varepsilon}^{2}%
{\mathrm{ln}{\left(  \frac{l-{\tilde{x}}_{1}}{l-{\hat{x}}_{1}}\right)  }}%
\end{equation}
and ${\mathrm{sin}\zeta\simeq\varepsilon\mathrm{ln}{\left(  \frac{l-{\tilde
{x}}_{1}}{l-{\hat{x}}_{1}}\right)  }}$. Using this, we deduce that the
numerator in the integrand will only have a small imaginary part and the real
part will be close to unity, namely
\[
\int_{0}^{l}{\frac{\left(  {\mathrm{cos}\zeta}\right)  d{\tilde{x}}_{1}%
}{{\left(  l-{\tilde{x}}_{1}\right)  }^{\frac{1}{2}}}\simeq}\int_{0}^{l}%
{\frac{d{\tilde{x}}_{1}}{{\left(  l-{\tilde{x}}_{1}\right)  }^{\frac{1}{2}}%
}-\varepsilon}^{2}\int_{0}^{l}{\frac{{\mathrm{ln}{\left(  \frac{l-{\tilde{x}%
}_{1}}{l-{\hat{x}}_{1}}\right)  }}d{\tilde{x}}_{1}}{{\left(  l-{\tilde{x}}%
_{1}\right)  }^{\frac{1}{2}}}}%
\]

Hence,

\noindent%
\begin{equation}
{\delta}_{2}\left(  {\hat{x}}_{1}\right)  =-\frac{\left(  c_{1}+c_{2}\right)
{\sigma}_{c}}{2\pi}\int_{{\hat{x}}_{1}}^{b}{\frac{{\hat{l}}^{\frac{1}{2}%
}\left[  1+2\varepsilon^{2}\left(  -2+\ln\left(  \frac{{\hat{l}}}{{\hat
{l}-\hat{x}}_{1}}\right)  \right)  \right]  }{{\left(  \hat{l}-{\hat{x}}%
_{1}\right)  }^{\frac{1}{2}}}}d\hat{l}\label{XX1}%
\end{equation}

\noindent

\noindent With $\varepsilon$ small, we retain only the first term not
depending on $\varepsilon^{2}$, and integrate to obtain

\noindent%
\begin{equation}
{\delta}_{2}\left(  {\hat{x}}_{1}\right)  \simeq-\frac{\left(  c_{1}%
+c_{2}\right)  {\sigma}_{c}}{2\pi}b\left[  \sqrt{\frac{b-{\hat{x}}_{1}}{b}%
}+\frac{{\hat{x}}_{1}}{2b}{\mathrm{ln}\frac{1+\sqrt{\frac{b-{\hat{x}}_{1}}{b}%
}}{1-\sqrt{\frac{b-{\hat{x}}_{1}}{b}}}\ }\right]
\end{equation}

\noindent

\noindent Now introduce $x=b-{\hat{x}}_{1}$ as the magnitude of the distance
from the tip of the cohesive zone. The result above then becomes%
\begin{equation}
{\delta}_{2}\left(  {x}\right)  \simeq-\frac{\left(  c_{1}+c_{2}\right)
{\sigma}_{c}}{2\pi}b\left[  \sqrt{\frac{x}{b}}+\frac{{1}}{2}\left(  1-\frac
{x}{b}\right)  {\mathrm{ln}\frac{1+\sqrt{\frac{x}{b}}}{1-\sqrt{\frac{x}{b}}%
}\ }\right]  \label{XX}%
\end{equation}

Based on the same deductions, the applied stress intensity factor causes a
cohesive zone stretch given approximately by

\noindent%
\begin{equation}
{\delta}_{A}\left(  x\right)  =\left(  c_{1}+c_{2}\right)  {K}_{A}\sqrt
{\frac{x}{8\pi}}=\frac{\left(  c_{1}+c_{2}\right)  \sigma_{c}b}{\pi}%
\sqrt{\frac{x}{b}} \label{YY}%
\end{equation}
where eqt.39 has been used to eliminate $K_{A}$ in favour of $b$.\ The total
cohesive zone stretch is given by Eq. (\ref{XX}) added to Eq. (\ref{YY}),
leading to%
\begin{equation}
{\delta}_{2,tot}\left(  {x}\right)  \simeq\frac{\left(  c_{1}+c_{2}\right)
{\sigma}_{c}}{2\pi}b\left[  \sqrt{\frac{x}{b}}-\frac{{1}}{2}\left(  1-\frac
{x}{b}\right)  {\mathrm{ln}\frac{1+\sqrt{\frac{x}{b}}}{1-\sqrt{\frac{x}{b}}%
}\ }\right]
\end{equation}

\noindent

\noindent as assumed for our calculations (\ref{barenblatt}). In retaining the
second order term,
\begin{equation}
{\delta}_{2,tot}^{2nd}\left(  {x}\right)  \simeq\frac{\left(  c_{1}%
+c_{2}\right)  {\sigma}_{c}}{2\pi}b\left[  \sqrt{\frac{x}{b}}-\frac{{1}}%
{2}\left(  1-\frac{x}{b}\right)  {\mathrm{ln}\frac{1+\sqrt{\frac{x}{b}}%
}{1-\sqrt{\frac{x}{b}}}-}\frac{\varepsilon^{2}}{b}\int_{{\hat{x}}_{1}}%
^{b}{\frac{2{\hat{l}}^{\frac{1}{2}}\left(  -2+\ln\left(  \frac{{\hat{l}}%
}{{\hat{l}-\hat{x}}_{1}}\right)  \right)  }{{\left(  \hat{l}-{\hat{x}}%
_{1}\right)  }^{\frac{1}{2}}}}d\hat{l}\right]
\end{equation}

Note that $c_{1}$ gives the contribution due to the upper material and $c_{2}$
that of the lower material.

Fig.4 shows the cohesive zone stretch due to the cohesive zone tractions only
without the cohesive zone stretch due to the applied stress intensity factor
i.e. from Eq.(\ref{XX1}) made dimensionless as $\widehat{{\delta}}_{2}\left(
{x}\right)  \simeq\delta_{2}\left(  {x}\right)  /\left(  -\frac{\left(
c_{1}+c_{2}\right)  {\sigma}_{c}}{2\pi}b\right)  $ for $\varepsilon=0$, (black
thick solid line), and for $\varepsilon=0.1,0.2,0.3$ to second order with
blue, red and green lines. While the slope of the black line should be
vertical at $x/b$ the effect is very confined so that it is not apparent on
the scale of Fig. 4. Notice that it can be shown by some lengthly algebra the
stretch of the crack outside the cohesive zone can be found by considering
absolute values of the quantities in the log terms, so we are plotting this
result too in Fig.4.

\begin{center}
\bigskip%
\begin{tabular}
[c]{l}%
\centering\includegraphics[height=65mm]{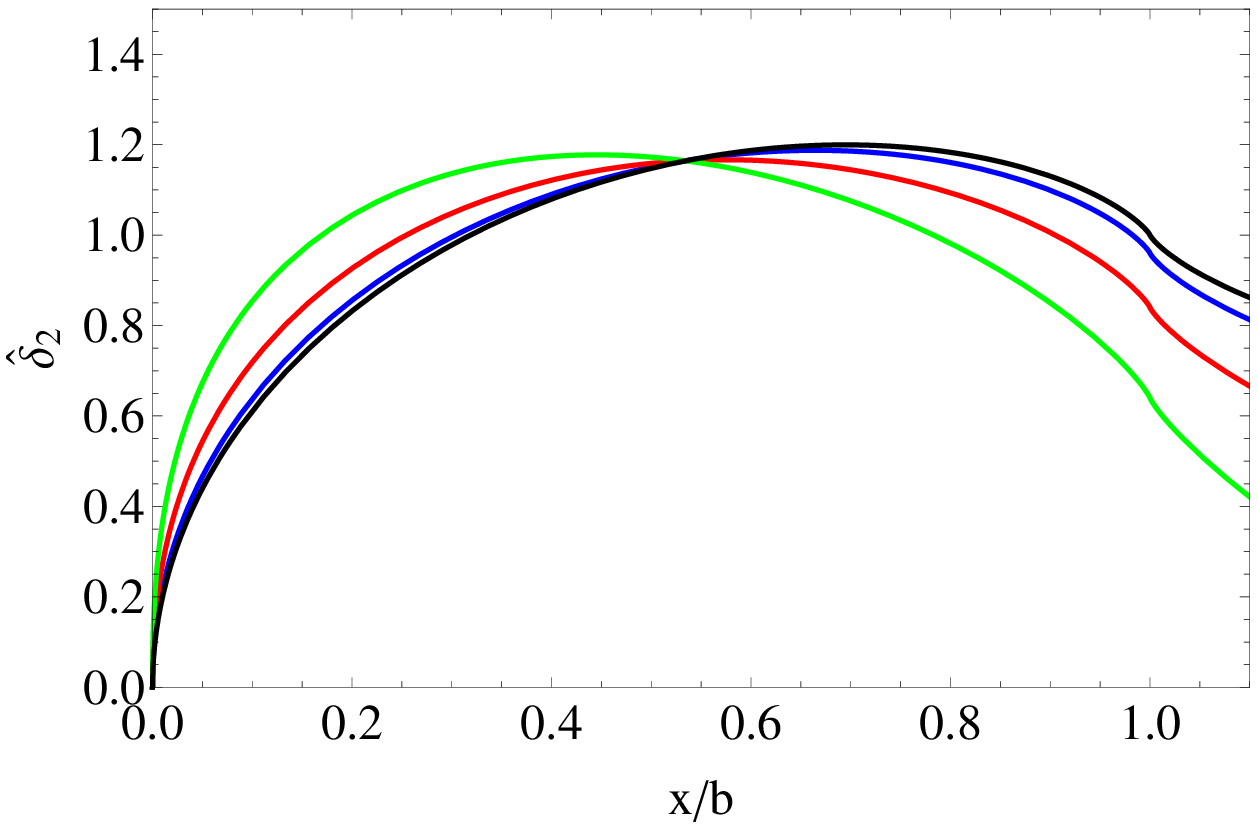}
\end{tabular}

Figure 4 -  Stretch due to the cohesive zone tractions only from
Eq.(\ref{XX1}) for $\varepsilon=0$, (black thick solid line), and for
$\varepsilon=0.1,0.2,0.3$ to second order with blue, red and green lines. 

\bigskip
\end{center}

\section{Appendix 2 - Derivation of cohesive zone stretch result with
viscoelasticity}

\noindent

\noindent The viscoelastic constitutive law is given by

\noindent%
\begin{equation}
{\varepsilon}_{ij}\left(  t\right)  =\int_{-\infty}^{t}{C_{ijkl}\left(
t-\tau\right)  \frac{d{\sigma}_{kl}\left(  \tau\right)  }{dt}d\tau}
\label{A21}%
\end{equation}

\noindent\noindent where ${\varepsilon}_{ij}$ is the strain, \textit{t} is
time, $C_{ijkl}\left(  t\right)  $ is the viscoelastic compliance and
${\sigma}_{kl}$ is the stress. Within a homogeneous material, Eq. (\ref{A21})
may be rewritten as

\noindent%
\begin{equation}
{\varepsilon}_{ij}\left(  t\right)  =\int_{-\infty}^{t}{C_{ijkl}\left(
t-\tau\right)  C_{klmn}^{-1}\left(  0\right)  \frac{d{\varepsilon}_{mn}%
^{el}\left(  \tau\right)  }{dt}d\tau}%
\end{equation}

\noindent where $C_{klmn}^{-1}\left(  0\right)  $ is the tensor of
instantaneous elastic moduli and ${\varepsilon}_{mn}^{el}\left(  \tau\right)
$ is the elastic strain at the current stress. We can integrate this with
respect to position in the homogeneous material to obtain

\noindent%
\begin{equation}
u_{i}\left(  t\right)  =\int_{-\infty}^{t}{\frac{C\left(  t-\tau\right)
}{C\left(  0\right)  }\frac{du_{i}^{el}\left(  \tau\right)  }{dt}d\tau}
\label{A23}%
\end{equation}

\noindent where $C\left(  t-\tau\right)  $ is an appropriate measure of the
viscoelastic compliance and $C\left(  0\right)  $ is the equivalent elastic compliance.

\noindent

\noindent Following Rice's (1978) treatment of the homogeneous case, we apply
Eq. (\ref{A23}) to each side of a crack along a bimaterial interface. We
consider incompressible materials, so that, after approximation of the
cohesive zone stretch, the contribution to the elastic stretch due to material
1 which is in the upper half of the plane, in a small cohesive zone on a
semi-infinite crack, is

\noindent%
\begin{equation}
{\delta}_{1}^{el}=\frac{{\sigma}_{c}b}{\pi G_{1}}\left[  \sqrt{\frac{x}{b}%
}-\frac{1}{2}\left(  1-\frac{x}{b}\right)  {\mathrm{ln}\frac{1+\sqrt{\frac
{x}{b}}}{1-\sqrt{\frac{x}{b}}}\ }\right]  =\frac{{\sigma}_{c}b}{\pi G_{1}%
}f\left(  \lambda\right)  \label{A24}%
\end{equation}

\noindent where $G_{1}$ is the shear modulus of material 1, $\lambda={x}/{b}$,
with an equivalent result for ${\delta}_{2}$, the contribution from material 2
in the lower half plane. In Eq.\ref{A24}, \textit{x} is measured from the tip
of the cohesive zone where ${\delta}_{1}^{el}=0$ and $x=b$ at the tip of the
crack where the cohesive zone will break. Thus, the length of the cohesive
zone is \textit{b} and we have, as an approximation,

\noindent%
\begin{equation}
b=\frac{\pi}{8}{\left(  \frac{K_{A}}{{\sigma}_{c}}\right)  }^{2}%
\end{equation}

\noindent

\noindent where $K_{A}$ is the applied Mode I stress intensity factor, used as
an approximation of the complex stress intensity factor for a bimaterial
crack, and ${\sigma}_{c}$ is the cohesive stress.

\noindent We define the viscoelastic compliance by

\noindent%
\begin{equation}
\gamma\left(  t\right)  =\int_{-\infty}^{t}{C\left(  t-\tau\right)
\frac{d{\sigma}_{s}\left(  \tau\right)  }{dt}d\tau}%
\end{equation}

\noindent where $\gamma$ is shear strain and ${\sigma}_{s}$ is shear stress.
Therefore, $C_{1}\left(  0\right)  ={1}/{G_{1}}$ and thus the viscoelastic
contribution of material 1 to the cohesive zone stretch is

\noindent%
\begin{equation}
{\delta}_{1}\left(  t\right)  =G_{1}\int_{-\infty}^{t}{C_{1}\left(
t-\tau\right)  \frac{d{\delta}_{1}^{el}\left(  \tau\right)  }{dt}d\tau}
\label{A27}%
\end{equation}

\noindent

\noindent Now consider a crack growing on the bimaterial interface, taken to
be the $x_{1}$-axis. The crack is growing at a rate \textit{V} such that its
tip, where the cohesive zone breaks, is at $x_{1}=Vt$ and the tip of the
cohesive zone is at $x_{1}=Vt+b$. As a result,

\noindent%
\begin{equation}
x=Vt+b-x_{1}%
\end{equation}

\noindent We consider steady state growth, with $K_{A}$ and \textit{b} both
constant, so that

\noindent%
\begin{equation}
\frac{d{\delta}_{1}^{el}\left(  x_{1},t\right)  }{dt}=\frac{{\sigma}_{c}b}{\pi
G_{1}}\frac{\partial}{\partial t}f\left(  \frac{Vt+b-x_{1}}{b}\right)
\end{equation}

\noindent As a result, Eq. (\ref{A27}) becomes

\noindent%
\begin{equation}
{\delta}_{1}\left(  x_{1},t\right)  =\frac{{\sigma}_{c}b}{\pi}\int_{-\infty
}^{t}{C_{1}\left(  t-\tau\right)  \frac{\partial}{\partial\tau}f\left(
\frac{V\tau+b-x_{1}}{b}\right)  d\tau}%
\end{equation}

\noindent To obtain the crack tip stretch due to material 1 we set $x_{1}=Vt$
so that

\noindent%
\begin{equation}
{\delta}_{1}\left(  Vt,t\right)  =\frac{{\sigma}_{c}b}{\pi}\int_{t-\frac{b}%
{V}}^{t}{C_{1}\left(  t-\tau\right)  \frac{\partial}{\partial\tau}f\left(
\frac{b-V\left(  t-\tau\right)  }{b}\right)  d\tau} \label{A28}%
\end{equation}

\noindent

\noindent where the lower limit on the integration arises as that is the time
material point that is the crack tip at time \textit{t} entered the cohesive
zone. We note that

\noindent%
\begin{equation}
\lambda=\frac{b-V\left(  t-\tau\right)  }{b}%
\end{equation}

\noindent

\noindent Thus, we can write

\noindent%
\begin{equation}
\tau-t=\frac{b}{V}\left(  \lambda-1\right)
\end{equation}

\noindent

\noindent enabling a change of the integration variable so that Eq.
(\ref{A28}) becomes

\noindent%
\begin{equation}
{\delta}_{1}^{tip}=\frac{{\sigma}_{c}b}{\pi}\int_{0}^{1}{C_{1}\left(  \frac
{b}{V}\left(  \lambda-1\right)  \right)  \frac{df\left(  \lambda\right)
}{d\lambda}d\lambda}%
\end{equation}

\noindent

\noindent where, as above,

\noindent%
\begin{equation}
f\left(  \lambda\right)  =\sqrt{\lambda}-\frac{1}{2}\left(  1-\lambda\right)
{\mathrm{ln}\frac{1+\sqrt{\lambda}}{1-\sqrt{\lambda}}\ }%
\end{equation}

\noindent

\noindent and thus

\noindent%
\begin{equation}
\frac{df\left(  \lambda\right)  }{d\lambda}=\frac{1}{2}{\mathrm{ln}%
\frac{1+\sqrt{\lambda}}{1-\sqrt{\lambda}}\ }%
\end{equation}

\noindent

\noindent The equivalent result for material 2 is

\noindent%
\begin{equation}
{\delta}_{2}^{tip}=\frac{{\sigma}_{c}b}{\pi}\int_{0}^{1}{C_{2}\left(  \frac
{b}{V}\left(  \lambda-1\right)  \right)  \frac{df\left(  \lambda\right)
}{d\lambda}d\lambda}%
\end{equation}

\noindent

\noindent Note that for a Maxwell material

\noindent%
\begin{equation}
C\left(  t\right)  =\frac{1}{G}+\frac{t}{\mu}%
\end{equation}

\noindent

\noindent where $\mu$ is the viscosity. Consequently,

\noindent%
\begin{equation}
\int_{0}^{1}{C_{1}\left(  \frac{b}{V}\left(  \lambda-1\right)  \right)
\frac{df\left(  \lambda\right)  }{d\lambda}d\lambda}=\frac{1}{G_{1}}+\frac
{b}{3{\mu}_{1}V}%
\end{equation}

\noindent It follows that the crack tip stretch for a Maxwell bimaterial crack
where the materials are incompressible is

\noindent%
\begin{equation}
{\delta}^{tip}={\delta}_{1}^{tip}+{\delta}_{2}^{tip}=\left[  \frac{1}{G_{1}%
}+\frac{1}{G_{2}}+\frac{b}{3V}\left(  \frac{1}{{\mu}_{1}}+\frac{1}{{\mu}_{2}%
}\right)  \right]  \frac{{\sigma}_{c}b}{\pi}%
\end{equation}

\noindent

\noindent
\end{document}